\documentclass[11pt,twoside]{article}
\usepackage{asp2014}

\aspSuppressVolSlug
\resetcounters

\begin{document}

\title{An introduction to FITSWebQL}

\author{C.~Zapart,$^1$ Y.~Shirasaki,$^1$ M.~Ohishi,$^1$ Y.~Mizumoto,$^1$ W.~Kawasaki,$^1$ T.~Kobayashi,$^1$ G.~Kosugi,$^1$ E.~Morita,$^1$ A.~Yoshino,$^1$ and S.~Eguchi$^2$
  \affil{$^1$National Astronomical Observatory of Japan, 2-21-1 Osawa, Mitaka, Tokyo 181-8588, Japan; \email{chris.zapart@nao.ac.jp}}
  \affil{$^2$Fukuoka University, 8-19-1 Nanakuma, Jonan-ku, Fukuoka 814-0180, Japan}}           

\paperauthor{C.~Zapart}{chris.zapart@nao.ac.jp}{}{National Astronomical Observatory of Japan}{Japanese Virtual Observatory}{Mitaka}{Tokyo}{181-8588}{Japan}
\paperauthor{Y.~Shirasaki}{}{}{National Astronomical Observatory of Japan}{}{Mitaka}{Tokyo}{181-8588}{Japan}
\paperauthor{M.~Ohishi}{}{}{National Astronomical Observatory of Japan}{}{Mitaka}{Tokyo}{181-8588}{Japan}
\paperauthor{Y.~Mizumoto}{}{}{National Astronomical Observatory of Japan}{}{Mitaka}{Tokyo}{181-8588}{Japan}
\paperauthor{W.~Kawasaki}{}{}{National Astronomical Observatory of Japan}{}{Mitaka}{Tokyo}{181-8588}{Japan}
\paperauthor{T.~Kobayashi}{}{}{National Astronomical Observatory of Japan}{}{Mitaka}{Tokyo}{181-8588}{Japan}
\paperauthor{G.~Kosugi}{}{}{National Astronomical Observatory of Japan}{}{Mitaka}{Tokyo}{181-8588}{Japan}
\paperauthor{E.~Morita}{}{}{National Astronomical Observatory of Japan}{}{Mitaka}{Tokyo}{181-8588}{Japan}
\paperauthor{A.~Yoshino}{}{}{National Astronomical Observatory of Japan}{}{Mitaka}{Tokyo}{181-8588}{Japan}
\paperauthor{S.~Eguchi}{}{}{Fukuoka University}{}{Nanakuma}{Fukuoka}{814-0180}{Japan}

  
  
\begin{abstract}

The JVO ALMA WebQL web service - available through the JVO ALMA FITS archive - has been upgraded to include legacy data from other telescopes, for example Nobeyama NRO45M in Japan. The updated server software has been renamed FITSWebQL. In addition, a standalone desktop version supporting Linux, macOS and Windows 10 Linux Subsystem (Bash on Windows) is also available for download from \url{http://jvo.nao.ac.jp/~chris/} .

The FITSWebQL server enables viewing of even 100GB-large FITS files in a web browser running on a PC with a limited amount of RAM. Users can interactively zoom-in to selected areas of interest with the corresponding frequency spectrum being calculated on the server in near real-time. The client (a browser) is a JavaScript application built on WebSockets, HTML5, WebGL and SVG.

There are many challenges when providing a web browser-based real-time FITS data cube preview service over high-latency low-bandwidth network connections. The upgraded version tries to overcome the latency issue by predicting user mouse movements with a Kalman Filter in order to speculatively deliver the real-time spectrum data at a point where the user is likely to be looking at. The new version also allows one to view multiple FITS files simultaneously in an RGB composite mode (NRO45M FUGIN only), where each dataset is assigned one RGB channel to form a colour image. Spectra from multiple FITS cubes are shown together too.

The paper briefly describes main features of FITSWebQL. We also touch on some of the recent developments, such as an experimental switch from C/C++ to Rust (see \url{https://www.rust-lang.org/}) for improved stability, better memory management and fearless concurrency, or attempts to display FITS data cubes in the form of interactive on-demand video streams in a web browser.
  
\end{abstract}

\section{Introduction}

Historically the ALMA WebQL service offered by the Japanese Virtual Observatory dates back at least to the year 2012 when it was presented during the ADASS XXII Conference\ \citep{2013ASPC..475..255E}. Afterwards, in order to keep up with ever growing FITS file sizes coming out of the ALMA observatory and also to offer improved functionality, newer versions have been released on a regular basis. For example, released in 2017, version 3 introduced an experimental 3D view of FITS data cubes. In 2018 the current version 4 --- completely re-written from scratch in the Rust programming language --- features real-time streaming videos of individual frequency channels in the FITS data cubes. The service can be accessed from the JVO Portal, found at \url{https://jvo.nao.ac.jp/portal/top-page.do} . The latest version 4 of the software (which includes the standalone desktop edition) is freely available from the following GitHub repository: \url{https://github.com/jvo203/fits\_web\_ql} . The unchanging motivation behind this web service is to provide a FITS file preview (quick look) and cut-out capability through a web browser. The service allows end users to view over 100GB-large FITS files in a web browser without ever having to download the underlying FITS files. After previewing FITS files users may choose to download interesting FITS files either in whole or to stream a partial region-of-interest (cut-out) from the JVO server to their own computers.

\section{Architecture}

The new version 4 initially started as a small feasibility study to find how easy it would be to re-implement the server part of FITSWebQLv3 in Rust. There are good reasons for switching from C/C++ to a new systems programming language such as Rust as it brings important benefits such as memory safety (\emph{no memory leaks}), thread safety (\emph{no data races}), better (smoother) multithreading compared with OpenMP in C/C++ and a complete lack of segmentation faults (\emph{no crashes}) due to inherent safety measures built into the Rust language. It is certainly possible to write C/C++ programs that are free of memory leaks and do not crash but from a programmer's standpoint Rust makes accomplishing these tasks much easier, all without sacrificing performance. In addition, Rust has an integrated HTTP/WebSockets networking library: {\texttt{actix-web}} that compares favourably with the previously used disparate mix of C libmicrohttpd and C++ $\mu$WebSockets.

With the Rust port under-way work had also been progressing on adding streaming video capability to the main v3 C/C++ codebase. However, once all the bottlenecks with Rust have been identified and dealt with, another benefit has come to light: the original C/C++ codebase has become rather complex and adding new functionality has turned into an error-prone process running the risk of introducing memory leaks and bugs. Hence a decision has been taken to complete the switch from C/C++ to Rust and add streaming video functionality to the new version v4, of which the full client-server architecture can be seen in Figures \ref{fig1}- \ref{fig2}.
\articlefigure{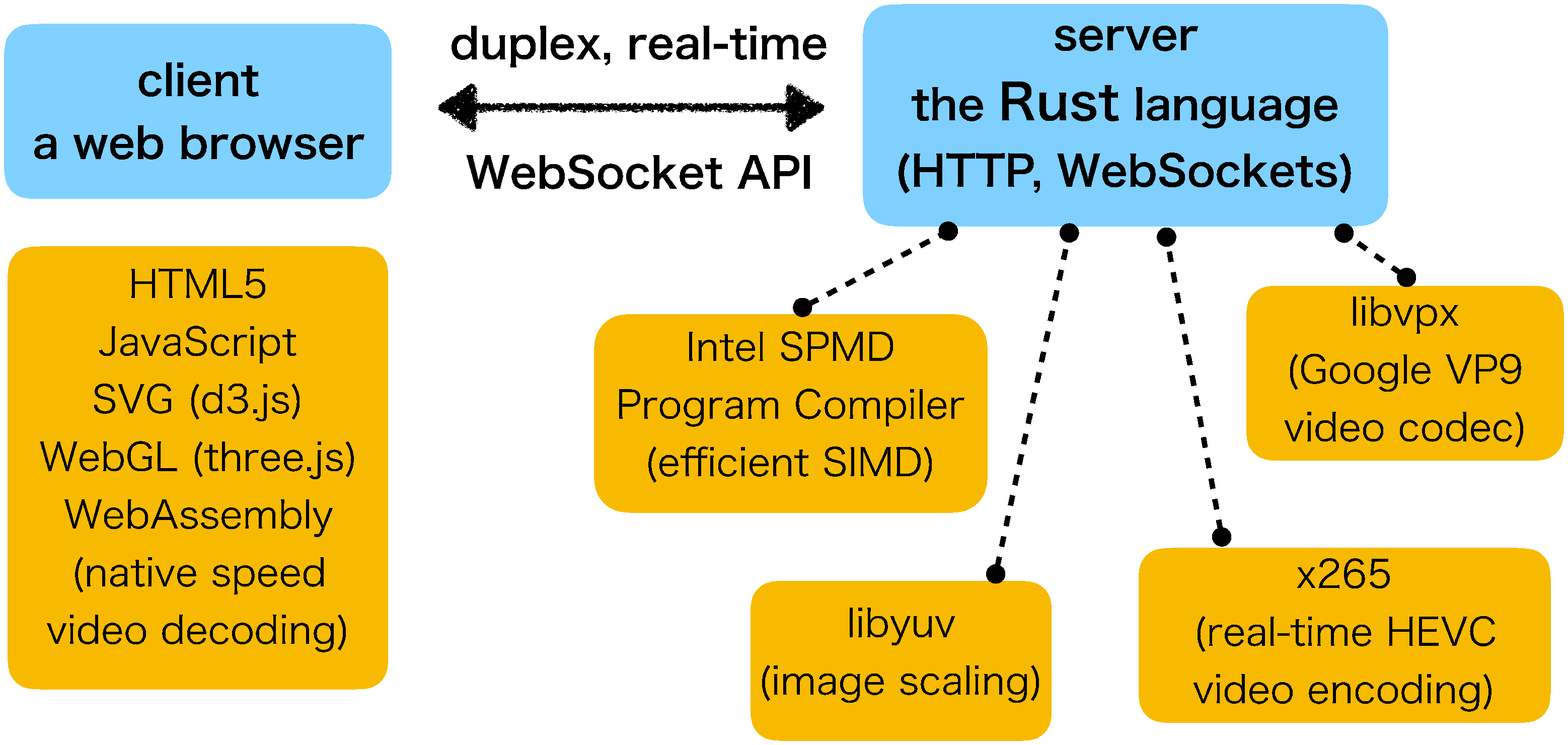}{fig1}{FITSWebQLv4 client-server architecture.}
\articlefigure{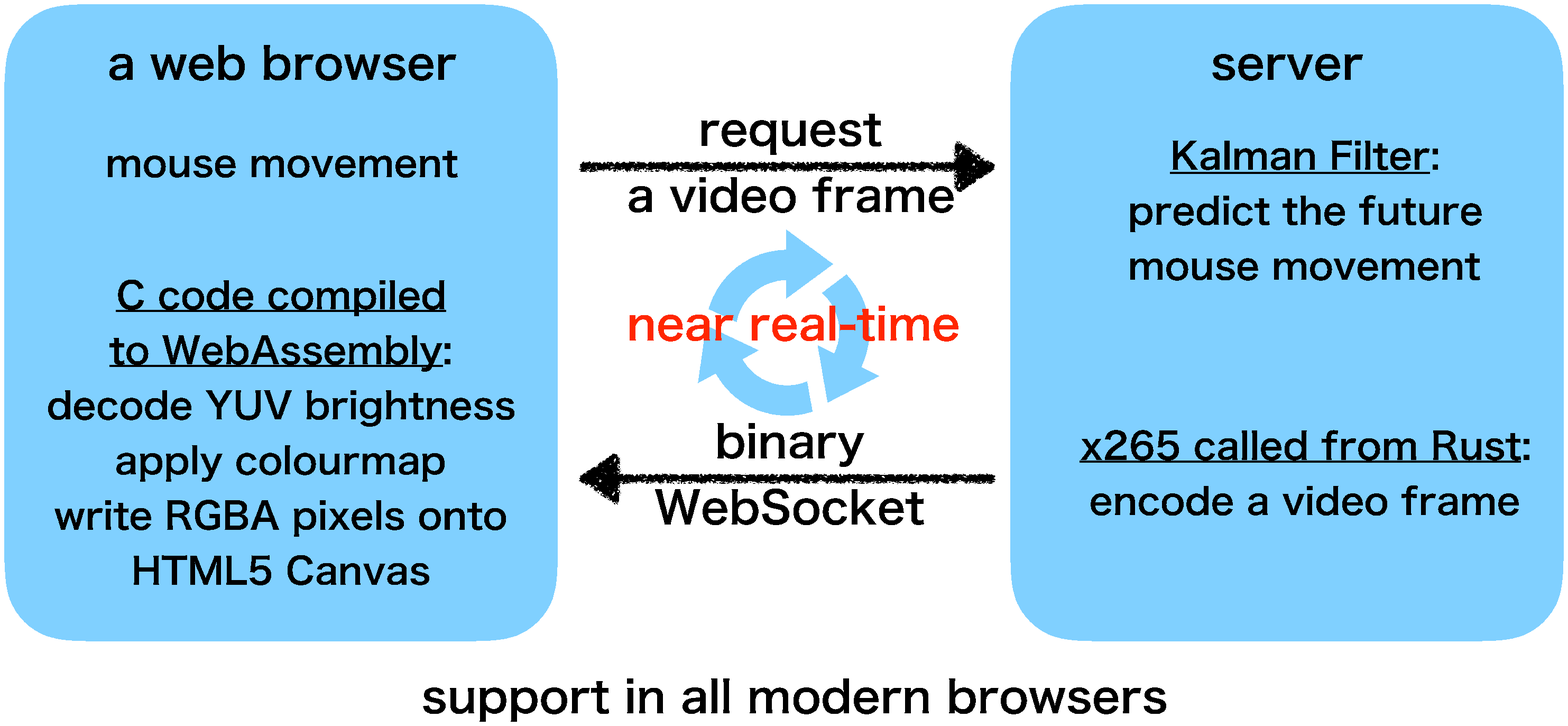}{fig2}{WebAssembly (Wasm) acceleration (near-native speed execution).}

A two-way communication between the client (a web browser) and the Rust server occurs over WebSockets, which halve the network latency and are more efficient in handling small messages compared to traditional AJAX HTTP requests. On the server side, the Rust language binds together various C/C++ libraries for which there is no high-performance 100\% pure-Rust implementation available. In particular, the computation-intensive parts are SIMD-parallelised using the Intel SPMD Program Compiler.\footnote{The open-source Intel SPMD compiler (see \url{https://ispc.github.io}) should not be confused with the paid-for Intel C/C++/Fortran compiler suite.} Unfortunately the \emph{no-crash} guarantees do not extend to external non-Rust libraries which may leak memory and may contain segmentation fault-causing bugs. One needs to be very careful when choosing which C/C++ libraries to call from Rust. 

On the client side we have taken advantage of the latest developments in browser technologies: the widespread adoption of WebAssembly (Wasm) that allows developers to compile C/C++ code to a binary Wasm stack machine code executed at a near-native speed inside a web browser.\footnote{\url{https://webassembly.org}} In particular the CPU-intensive parts: real-time decoding of HEVC video frames and the application of a user-specified colourmap to greyscale video frames have been greatly accelerated with WebAssembly.

\subsection{VP9 vs. HEVC comparison}

During the initial development originally the author intended to use the HEVC (via its x265 encoder library) codec to handle real-time video streams. However, finding a suitable JavaScript and/or Wasm decoder has proved impossible. The resources freely available on the Internet did not meet our requirements. They were too outdated; they did not support the latest HEVC specification. As an alternative, after exploring other codecs i.e. Cisco's Thor, initially we integrated Google's VP9 libvpx library into our project. However, due to inferior multithreading capabilities of libvpx and codec inefficiencies compared with a superior HEVC solution, we decided to return to using HEVC/x265. Since there was no suitable off-the-shelf HEVC browser decoder available, we were forced to adapt the HEVC decoding part from the FFmpeg C library and compile it to WebAssembly for fast native execution in a web browser.

As a result of this somewhat convoluted development process, as of now the VP9 library is still used to compress FITS 2D images (as VP9 still keyframes) for display in a browser whilst the more capable HEVC x265 library handles real-time video streaming. Table \ref{tab1} shows the main pros and cons of the two codec formats.
\begin{table}[!ht]
\caption{A side-by-side comparison of Google's VP9 and HEVC video codecs together with their corresponding C API libraries.}
\label{tab1}
\smallskip
\begin{center}
{\small
\begin{tabular}{ll}  
\tableline
\noalign{\smallskip}
Google's VP9 (libvpx) & HEVC (x265) \\
\noalign{\smallskip}
\tableline
\noalign{\smallskip}
libvpx library: both an encoder and decoder & x265 library: only an encoder \\
& (search the Internet for a decoder \\
& to suit your task) \\
\noalign{\smallskip}
\tableline
\noalign{\smallskip}
slower, less efficient encoding, & faster than libvpx, more efficient \\
inferior multithreading & (bandwidth-friendly), scales well \\
& across all CPU cores \\
\noalign{\smallskip}
\tableline
\noalign{\smallskip}
no greyscale (an overhead of handling & YUV 4:0:0 support (server-encode as \\
redundant RGB/YUV channels) & greyscale, add colour in the client) \\
\noalign{\smallskip}
\tableline
\noalign{\smallskip}
an easy API, trivial to compile the & extreme difficulty finding a suitable \\
decoder into WebAssembly & JavaScript decoder (DIY: FFmpeg C \\
& API compiled to WebAssembly)\\
\noalign{\smallskip}
\tableline\
\end{tabular}
}
\end{center}
\end{table}

\section{Conclusions}

Based on our experience at JVO Rust has largely lived up to its promises. Not only did performance not deteriorate, in a few places it has actually improved compared to the C/C++.
The only disadvantage of Rust seems to be its steep learning curve.  

\section{References}

\bibliography{O3-1}

\end{document}